**Nanocomposites of ferrocenyl-modified gold clusters and semiconducting polymers that integrate field-effect transistor and flash memory in a single neuromorphic device**

Deepa Singh[1], Praveen N. Gunawardene[2], Mark S. Workentin[2], and Giovanni Fanchini[1,2,] *

[1]*Department of Physics & Astronomy, The University of Western Ontario, 1151 Richmond St., London, Ontario, N6A 3K7, Canada.*

[2]*Department of Chemistry, The University of Western Ontario, 1151 Richmond St., London, Ontario, N6A 5B7, Canada.*

**Abstract**

We demonstrate that electroactive thin films incorporating semiconducting polymers and deterministic functionalized gold nanoclusters (ncAu$_{25}$) lead to integration of the functions of resistive memory device and field-effect transistor (FETs) within a single component ("mem-transistor") in a neuromorphic system. Memristor functions originate from ferrocenyl-modified gold nanoclusters (ncAu$_{25}$-Fc) embedded in polymethyl-methacrylate (PMMA) and devices optimized for maximum 1/0 "flash" memory effect are found to contain 15 wt% ncAu$_{25}$-Fc. Integrated memristor and neuromorphic functions are obtained by replacing PMMA with poly(3-hexylthiophene) (P3HT) in the active layer, from which transistor effects are derived. Based on the energy band diagrams of ncAu$_{25}$, PMMA and P3HT, percolation theory is used to explain the memristor 1/0 on/off ratio as a function of ncAu$_{25}$-Fc concentration. The use of ncAu$_{25}$-Fc with charge-tunable, ferrocene-modified, ligands is critical toward better cluster-polymer interfaces. Our work shows that nanostructures of polymers and metalorganic frameworks bear strong potential towards neuromorphic devices and the circuital simplification of data storage technology.

______________________________

* *Corresponding author. E-mail*: gfanchin@uwo.ca

Flash memory devices capable of performing read, write, and erase operations multiple times are essential components in information technology.[1] Commercial inorganic flash memory devices are normally designed with one-capacitor/one-transistor (1C/1T) type architectures, in which the capacitor operates as a data-storage element while a field-effect transistor (FET) is used as a switch to access the stored data.[1] Inorganic mem-transistors combining the functions of resistive memory devices and field-effect transistors within the same component have recently been reported,[2,3] but they require high-temperature (800°C) fabrication processes that are not compatible with the needs of low-cost and flexible electronics[4,5]. Although organic FETs are most suited for the realization of flexible and cost-efficient electronics, they are not multifunctional,[6,7] and nanostructured hybrid systems have been proposed as alternatives.[8] Particularly attractive are neuromorphic devices, in which transistor and memory effects are integrated within the same component like in cerebral and artificial neural networks.[9]

Molecular gold clusters (ncAu$_{25}$) containing a deterministic number of Au atoms (e.g., 25, 38, or 144) with diameters below 2 nm have been widely investigated due to their deterministic size-related properties, simplistic surface modification, and high thermal stability.[10,12,13] ncAu$_{25}$ serve as charge storage units within the polymer matrix enabling them to function as a memristor.[14] There are a few reports of electronic devices based on gold nanocluster-polymer nanocomposites,[14,] but none of them has demonstrated gating, field-effect, and memristive effects through the ability of gold nanoclusters to act as dopant–a phenomenon that, as far as organic electronics is concerned, has been so far confined to ionic nanocluster-evolved polymers.[15]

Here, we demonstrate that ncAu$_{25}$ modified with ferrocene-containing ligands and embedded in a suitable polymer matrix can function in FETs that are integrating field effects and resistive flash memory effects in a single device, thus effectively operating as "memtransistors".

Thiolate-protected ncAu$_{25}$ have been shown to exhibit resistive memory effects, but such effects are due to their Au core.[14] Similar to Au$_{25}$ molecular cores, ferrocene-containing ligands also possess tunable charge states, "0" and "1", due to double oxidation levels.[16] In this work ferrocenyl-modified ncAu$_{25}$ (here referred as ncAu$_{25}$-Fc for the sake of simplicity) were prepared from a cluster surface strain-promoted azide alkyne cycloaddition reaction between a ferrocenyl ester of bicyclo[6.1.0]non-4-yne, which are combined with [Au$_{25}$(SCH$_2$CH$_2$-p-C$_6$H$_4$-N$_3$)$_{18}$]$^-$ to yield a cluster with up to 18 ferrocene units per ncAu$_{25}$.[17,18] The synthesis procedure is described in detail by Gunawardene et al.[17] The characterization procedure to ensure successfulness of the synthesis process includes proton nuclear magnetic resonance, near ultraviolet-visible (UV-vis) and infrared spectroscopy as well as matrix-assisted laser desorption/ionization mass spectrometry and is reported in detail in Ref. 18. It is anticipated that the extension of tunable electronic orbitals and tunable oxidation states to the outer shell would increase the ability of ligand-protected ncAu$_{25}$. Thus, the addition of ferrocene would affect the electronic properties of the environment in which they are embedded, thus producing mem-transistor effects.

To explore the role of ferrocenyl-modified ncAu$_{25}$ clusters in polymer matrixes, we fabricated two sets of electronic devices: (i) memristors with metal/insulator-ncAu$_{25}$-Fc/metal architecture, and (ii) FETs with ncAu$_{25}$ embedded in the active layer. For memristor fabrication, a solution of ncAu$_{25}$/PMMA in chlorobenzene was used. ncAu$_{25}$ were dispersed in 1 ml of chlorobenzene via ultrasonication for 30 minutes. The amount of ncAu$_{25}$ was varied as 0, 1.5, and 3 mg. The final amount, after adding poly(methyl methacrylate) (PMMA), of the solute material (PMMA+ ncAu$_{25}$) was fixed to be 10 mg. The solution was then stirred for 6 hours. The resulting solution was spun on cleaned ITO coated glass substrate with a spin coater (WS 400-6NPP, Laurell Technologies Co.). The spin speed was 2000 RPM for 1 minute and the resultant film thickness was measured to be 50 nm. The PMMA-ncAu$_{25}$ composite films were annealed at 120 ˚C for 20 minutes, a temperature that was previously shown[13] to leave ncAu$_{25}$ intact while agglomerating them into globules. Aluminum was thermally evaporated onto the PMMA-ncAu$_{25}$ films as a top electrode (anode) as shown in Fig.1a.

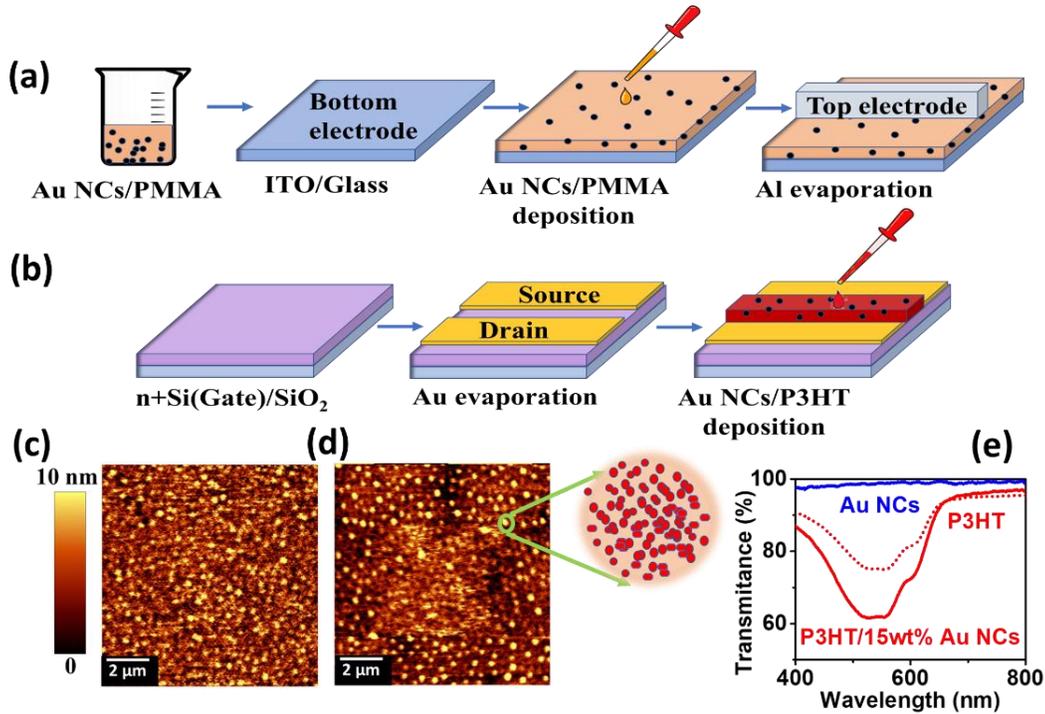

*Figure 1: Demonstration of fabrication of ncAu$_{25}$ based (a) memristors and (b) mem-transistors. AFM of ncAu$_{25}$-Fc (15 wt%) embedded in (c) PMMA and (d) P3HT. AFM images show uniformly distributed nanosized globules consisting of large groups of Au$_{25}$ nanoclusters. (e) UV-vis spectra of pure P3HT and P3HT / ncAu$_{25}$-Fc composite films used in this study. A decrease in P3HT transmittance is observed upon addition of ncAu$_{25}$. A UV-vis spectrum of a thin-film only made up by ncAu$_{25}$-Fc is shown as a reference.*

For the fabrication of FETs, Au source and drain electrodes were deposited on cleaned n$^+$Si/SiO$_2$(100 nm) substrate by a shadow mask as depicted in Figure 1b. For the semiconductor layer, Poly(3-hexylthiophene) (P3HT) with 15wt% of ncAu$_{25}$-Fc solution was spun on the n$^+$Si/SiO$_2$ substrate. The P3HT- ncAu$_{25}$ thin films were subjected to thermal annealing at 120˚C for 10 minutes to improve the crystallization of P3HT. The annealing temperature of P3HT-ncAu$_{25}$ composite films (120 ˚C) was chosen such that it would not affect the ligands in ncAu$_{25}$.[13] For surface morphology characterization of the films, atomic force microscopy (AFM) measurements were carried out by Witec Alpha 300S atomic force microscope system. UV−Vis spectra of spun films were recorded on a Varian DMS80 spectrophotometer.

Figure 1c and d depict the atomic force microscopy (AFM) images of PMMA and P3HT composite films with 15 wt% of ncAu$_{25}$-Fc respectively. Both films consisted globules of ncAu$_{25}$-Fc with variable size in the range of 100-150 nm. These globules were formed by the agglomeration of ncAu$_{25}$-Fc as also shown by Bauld et al.[13]. Although we have annealed the composite films, the annealing temperature is quite low and would not affect sulfur (S) atoms in Au$_{25}$ and S atoms will be intact in ncAu$_{25}$. As the S atoms are crucial for the stability of gold clusters,[19] ncAu$_{25}$-Fc will not lose their properties, not even as aggregates in the solid state.

We also investigated the effect of inclusion of ncAu$_{25}$-Fc in the polymer matrix via UV–Vis transmission spectroscopy as this is among the most popular techniques for the structural characterization of gold clusters.[19] Fig. 1e compares the UV-Vis spectra of P3HT with 15 wt% ncAu$_{25}$ composite films with pristine P3HT thin films. The spectral shape of the composite films was found to be very similar to the pristine P3HT films with a broad peak with a maximum at 520 nm corresponding to π-π* transitions of electrons. This indicates that the addition of ncAu$_{25}$-Fc did not affect the degree of crystalline order of P3HT.

To achieve the ideal concentration of ncAu$_{25}$ for memristive properties, we have investigated devices at different ncAu$_{25}$-Fc concentration and the resulting effects on the conduction behavior of ITO/PMMA/Al devices. We fabricated devices with PMMA containing 0,15 and 30 wt % ncAu$_{25}$-Fc concentration as an active layer. We measured the current (*I*)- voltage (*V*) response of these devices, by applying an external voltage with a sequence of 0 → 5 V → -5 V → 0 V and recording the resulting current. The current-voltage (*I-V*) scan of device comprising only PMMA is shown in figure 2a. Due to the insulating nature of PMMA, these device exhibits only a low conductivity state. Devices with a low number of nanoclusters (data not shown here) do not show any significant difference in electrical behavior.

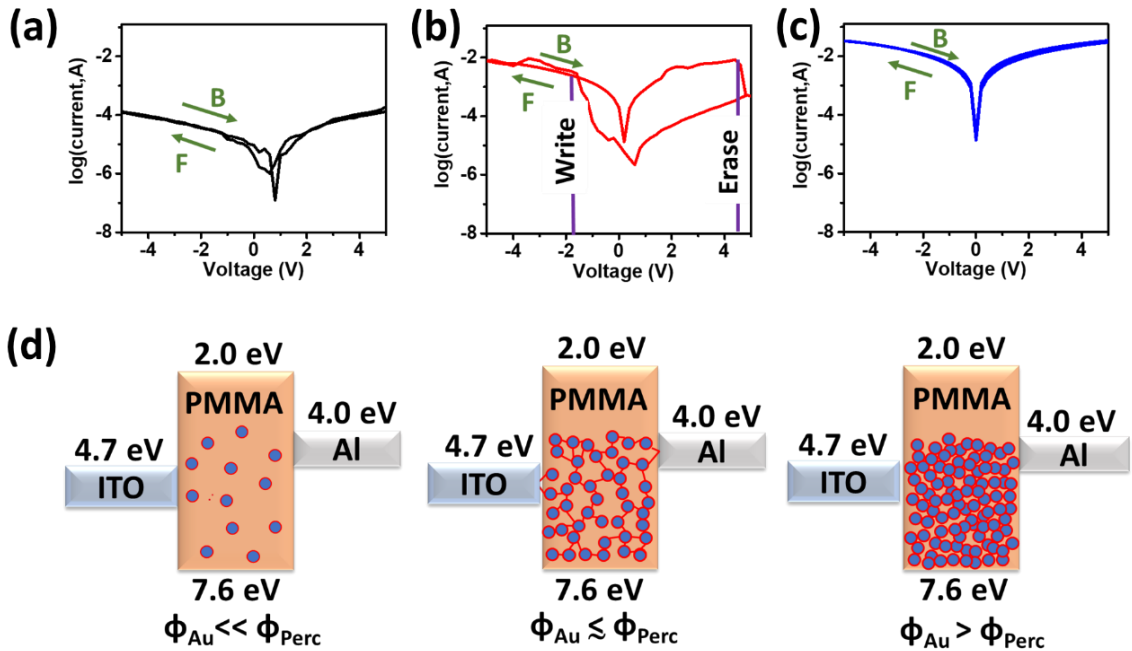

*Figure 2*: I-V characteristics of the ITO/PMMA- ncAu$_{25}$s/Al devices with (a) 0%, (b) 15% and (c) 30wt% ncAu$_{25}$-Fc. PMMA-15wt% ncAu$_{25}$-Fc composite films displayed bistable memory functionality with "write" and "erase" voltages are indicated. Green arrows along the I–V curves show the forward (F) and backward (B) sweep direction. Energy band diagrams in (d) represent the conduction mechanism in 0, 15 and 30 wt% ncAu$_{25}$-PMMA composite devices respectively.

Figure 2b shows the I-V characteristics of the devices with 15 wt% ncAu$_{25}$-Fc content. At $V = 0$ V, the devices were in low conductivity (0) state until a "write" voltage of about -1.7 V was attained. At this voltage, devices switched from low to high conductivity (1) state as the magnitude of current increased from 0.1 mA to 10 mA. The device remains in the same state until an "erase" voltage condition was reached at 4.5 V. At this voltage, the device switched again and returns to the initial state. Furthermore, the devices were in high conductivity state even when the power supply is switched off demonstrating the non-volatile nature of memory. Hence, these devices can function as non-volatile memory resistors. Figure 2c shows the electrical current behavior with respect to the applied voltage for the devices with 30 wt% Fc-ncAu$_{25}$ content. We did not observe any resistive switching in these devices as they were in a high conductivity state during the entire voltage cycle. The high conductivity is due to the formation of conductive networks of ncAu$_{25}$-Fc in PMMA films. These results indicate that distinct device behaviors namely insulators, nonvolatile bistable memristors, and conductors can be realized by increasing the nanocluster content in the PMMA/ ncAu$_{25}$ composite films.

We can explain the tunability of the electrical conductivity of these devices in terms of the electronic structure of ITO/PMMA-ncAu$_{25}$-Fc /Al systems. As the work function of Al (4 eV) is close to the lowest occupied molecular orbital level of PMMA (2 eV), a negative bias at Al contact will result in the injection of electrons into the composite film. These injected electrons will be occupied in nanocluster sites. When the number of nanocluster trapping sites is well below the percolation threshold,[20] the trap sites will be far from each other and hence the occupied electrons will also be isolated (Figure 2d). These trapped electrons will not contribute to the conduction leading to a low current. Even at higher voltages, these electrons will not be sufficiently mobile, and devices would remain in the low conductivity state.

As we further increase the ncAu$_{25}$-Fc concentration in PMMA films, the distance between the gold clusters will decrease (fig. 2e). At low bias, due to trapping of injected electrons, the conductivity of the devices will be low. As we increase the bias, the electrons will gain activation energy from the applied bias and their mobility will increase. Now the electrons would be able to hop from one site to another due to the proximity of nanocluster trapping centers. The inter-site hopping (depicted with the red lines between ncAu$_{25}$sites) will cause a sudden increase in current allowing devices to switch from low to high conductivity states. When a positive voltage is applied to Al contact, the trapped electrons in the ncAu$_{25}$-Fc sites will be neutralized by positive charges causing the device to return to its initial low conductivity state. Therefore, the trapping of electrons will cause a "0" state and the de-trapping of electrons would result in "1" state enabling the devices to function as a resistive memory. Figure 2f shows the energy band diagram when the concentration of nanoclusters is equal or above the percolation threshold. In this condition, ncAu$_{25}$-Fc sites are very compactly packed and provide a continuous path for charge carriers to effectively transport between the two electrodes. This will make devices highly conductive even at low bias.

The root cause for memory effects in optimized devices with 15 wt% ncAu$_{25}$-Fc content was further explored by investigating the charge transport process in these systems. A negative voltage at the Al contact lowers the Fermi level by an amount equal to the applied bias. Hence band bending occurs as shown in fig. 3a. When the applied voltage is less than the potential barrier height at the Al/PMMA interface (i.e., $V < q\varphi$), electrons can flow through the PMMA film via tunneling between the nanocluster traps, following a direct tunneling conduction mechanism. When $V < q\varphi$, more traps will be filled and more electrons would tunnel through a triangular barrier following Fowler–Nordheim (FN) tunneling. The trap assisted direct tunneling regime is shown by the linear dependence of $ln(I/V^2)$ on $ln(1/V)$[21] while FN tunneling regime is shown by the linear dependence of $ln(I/V^2)$ on $(1/V)$[21] as depicted in Figure 3b and c respectively. An inflection point in fig. 3c highlights the transition from direct to FN tunneling regime. This inflection point denotes the "write" voltage where the current will switch from low-conductivity to high-conductivity regime. Such a "write" voltage value (1.8 V) is in good agreement with previous estimates of a barrier height of ~2 eV at the Al/PMMA interface. Any further increment of voltage will cause the full occupation of trap sites by injected electrons. PMMA will be trap-free and injected carriers will be free to move in the system. Conduction becomes ohmic ($I \propto V$) as in region 3 of fig. 3d.

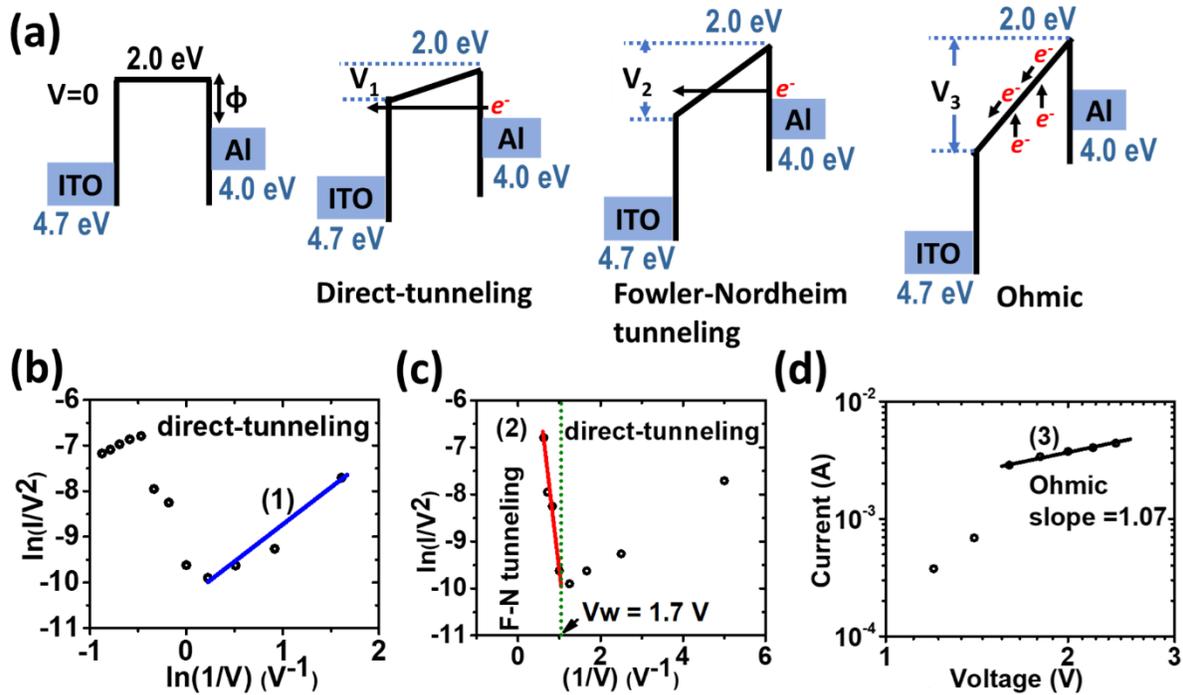

*Figure 3: (a) Schematic illustrations of the charge carrier transport and corresponding fitted I-V plots for (b) direct tunneling (c) Fowler-Nordheim tunneling and (C) ohmic conduction respectively. The curve in (c) also shows an inflection point denoted as write voltage ($V_w$) where devices flip from low to high conductivity state.*

We also fabricated FETs with P3HT-ncAu$_{25}$-composite films (15 wt% in ncAu$_{25}$-Fc) as an active layer. FETs without ncAu$_{25}$-Fc were also fabricated as control samples. Fig. 4(a–b) show the output curves obtained by measuring drain-source current ($I_{DS}$) with respect to the Drain-source voltage ($V_{DS}$). The FETs showed typical p-type transistor behavior[21] where $I_{DS}$ increased with gate voltage ($V_G$). The relatively low $I_{on}/I_{off}$ ratio of our FET devices is also common to our control (i.e., ncAu$_{25}$-Fc free) devices. The $I_{on}/I_{off}$ ratio can perhaps be improved by utilizing next-generation semiconducting polymers for organic electronics in lieu of commercial-grade P3HT. To investigate the charge storage in ncAu$_{25}$-Fc, $I_{DS}$-$V_{DS}$ curves were measured by applying forward (5 → -10V) and reverse bias (-10 → 5 V) cycles at a constant $V_G$ = -5V. Hysteresis is observed in $I_{DS}$-$V_{DS}$ plots in FETs with P3HT-ncAu$_{25}$-composite films, while lack of hysteresis demonstrates negligible memory effects in control devices without ncAu$_{25}$-Fc (fig. 4c).

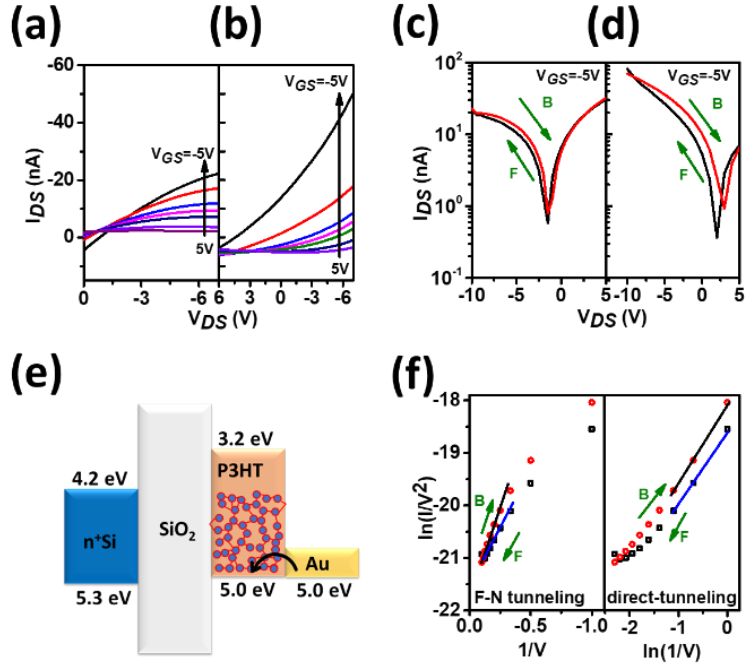

*Figure 4: Output characteristics of FETs with (a) P3HT and (b) P3HT-15wt% ncAu$_{25}$-Fc as an active layer. Log $I_{DS}$ -$V_{DS}$ plots for FETs with (c) P3HT and (d) P3HT with15wt% ncAu$_{25}$-Fc in response to a forward and backward voltage sweep. A mem-transistor behavior was shown by FETs having ncAu$_{25}$-Fc embedded in P3HT. (e) Energy band diagram of P3HT-15wt% ncAu$_{25}$-Fc based FETs. The bandgap of SiO$_2$ is 9.0 eV. (f) F-N and direct tunneling plots for memtransistors in the $V_{DS}$ range of 0 to -10V. Dotted lines indicate the linear parts of the curves.*

Fig. 4d presents the $I_{DS}$-$V_{DS}$ curve for FETs with 15 wt% ncAu$_{25}$-Fc. Transistors have significant hysteresis in their current response. The origin of the hysteresis lies in charge trapping capabilities of gold nanoclusters. Under positive voltage at Au electrode, holes are injected into the P3HT layer as the work function of Au (5.0 eV)[22] is close to the highest occupied molecular orbital level of P3HT (Fig. 4e). The holes will be occupied by Au nanoclusters trapping sites leading to a net hole trapping effect. When a negative bias is applied to Au electrode, the confined charges in the trap states will move back to P3HT films resulting in an increment of output current. Thus, charge trapping in ncAu$_{25}$-Fc will lead to a low-conductivity (0) state, while the release of charges from the trap sites will cause a high-conductivity (1) state to turn on.

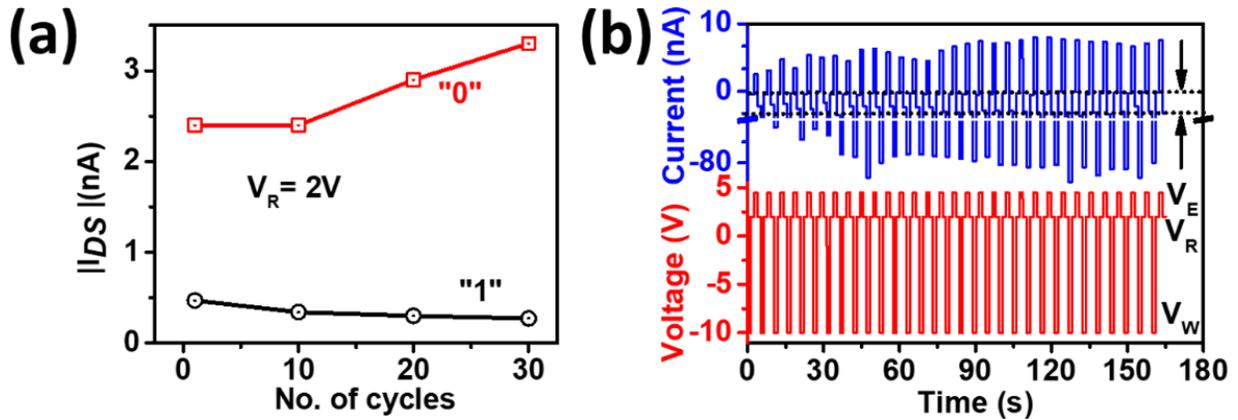

*Figure 5: (a) Reproducibility of memtransistors with P3HT and 15wt%. ncAu$_{25}$-Fc. Both "0" and "1" states were read at 2 V. No change in device performance observed even after 30 switching cycles. (b)The current response of the same memtransistors corresponding to write–read–erase–read voltage cycles. Top and bottom curves show the current response and corresponding applied voltage, respectively. FETs maintain their on/off ratios over ~30 switching cycles.*

Charge transport processes in memtransistors are found to be similar to their memristor counterparts. Injected holes from source reach the drain electrode via tunneling through cluster traps. At lower bias, the current is significantly low following direct tunneling model while at higher bias current will increase satisfying F-N tunneling model. In the backward sweep, the current follows the same trend but with higher slopes as evident in figure 4f. The large slope of the current-voltage scan signifies the increase in the amount of the current. This happens because in backward sweep, the holes will be released from the trap sites will now contribute to the current.

In addition to charge trapping inside the semiconducting film, metal/semiconductor interfaces can also capture the charge carriers[23]. In order to demonstrate that charge confinement in ncAu$_{25}$-Fc is the major factor of influence for observed memory effects in memtransistors, we examined their stability by applying multiple cycles in forward and backward directions. The "0" and "1" currents were measured at "read" voltage $V_R$ = 2 V in forward and backward sweeps of $V_{DS}$. The observed current is plotted as a function of the number of switching cycles in fig. 5a. Almost no degradation in was observed for over 30 voltage cycles. These results confirm that the charge trapping and release by ncAu$_{25}$ trap sites is the major cause of the memory effect in the mem-transistors in a neuromorphic device.[9] We have also performed reproducibility test on the memtransistors under repeated write-read-erase-read voltage cycles. The results are shown in Figure 5b. The current response of the mem-transistors [top curve in figure 5(b)] for the applied read-write-read-erase voltage sequence, indicates reproducible switching behavior.

In conclusion, we fabricated specific mem-transistors, in which flash memory effects were induced in an organic field-effect transistor via ferrocene-modified ncAu$_{25}$ molecular clusters. We optimized the memory effects of ncAu$_{25}$ into an ITO/PMMA-ncAu$_{25}$/Al architecture utilizing electrical conductivity measurements, in which optimum memory functionality was found to be at 15 wt% ncAu$_{25}$-Fc. With this optimal concentration in mind, we then demonstrated mem-transistors consisting of an ITO/PMMA- ncAu$_{25}$-Fc /Al architecture with 15 wt% ncAu$_{25}$-Fc in P3HT, which showed memory effects that are not present in P3HT field effect transistors. These mem-transistors displayed high stability and low voltage (< 10 V) operation. Neuromorphic computing mimics the approach of the human brain in integrating multiple functions of data storage and data processing within the same neuron-like component. Therefore, since our systems eliminate the need for a separate charge storage unit in 1C/1T memories, they can be incorporated into next-generation neuromorphic devices. Furthermore, our memristors can be easily made flexible, as they can be assembled on soft and plastic substrates. Therefore, owing to their integration capability in flexible electronics, they can be successfully interfaced with the human body and human brain.

## Acknowledgements

This work was supported by the Canada Research Chair program (GF), the Canada Foundation for Innovation (GF) and the Natural Sciences and Engineering Research Council of Canada under the discovery Grant program (MSW and GF).

## Conflicts of Interest

The authors have no conflicts to disclose.